\documentclass[twocolumn]{aastex6}

\usepackage{hyperref}

\usepackage{amsmath}

\shorttitle{Probing the innermost accretion flow geometry of Sgr A$^{*}$ with EHT}
\shortauthors{Pu \& Broderick}
\begin{document}


\title{Probing the innermost accretion flow geometry of Sgr A$^{*}$\\ with Event Horizon Telescope}
\author{Hung-Yi Pu\altaffilmark{1,2} } 
\author{Avery E. Broderick\altaffilmark{3,1}}

\altaffiltext{1}{Perimeter Institute for Theoretical Physics, 31 Caroline Street North, Waterloo, ON, N2L 2Y5, Canada}
\altaffiltext{2}{Institute of Astronomy \& Astrophysics, Academia Sinica, 11F of Astronomy-Mathematics Building, AS/NTU No. 1, Taipei 10617, Taiwan}
\altaffiltext{3}{Department of Physics and Astronomy, University of Waterloo, 200 University Avenue West, Waterloo, ON, N2L 3G1, Canada}

\begin{abstract}
Upcoming  Event Horizon Telescope (EHT) observations will provide a unique opportunity to reveal the innermost region of the radiative inefficient accretion flow (RIAF) around the Galactic black hole, Sgr A$^{*}$.
Depending on the flow dynamics and accumulated magnetic flux, the innermost region of an RIAF could have a quasi-spherical or disk-like geometry. 
Here we present a phenomenological model to investigate the characteristics of the black hole shadow images with  different flow geometries, together with the effect of black hole spin and flow dynamics.
The resulting image consists  in general of two major components: a crescent,  which may surround the funnel region of the black hole or the black hole itself, and a photon ring, which may be partially luminous and overlapped with the crescent component.
Compared to a quasi-spherical flow case, a  disk-like flow in the vicinity of a black hole exhibits  the following image features: (i) due to  less  material near the funnel region, the crescent structure  has a smaller size, and (ii) due to the  combination of emission from the flow  beside and behind the black hole, the crescent structure has a more irregular shape, and a less smooth brightness distribution. 
 How these features can result in different observables for EHT observations is discussed.

\end{abstract}


\section{Introduction}

\begin{figure*}
  \begin{center}
    \includegraphics[angle=0,width=\textwidth]{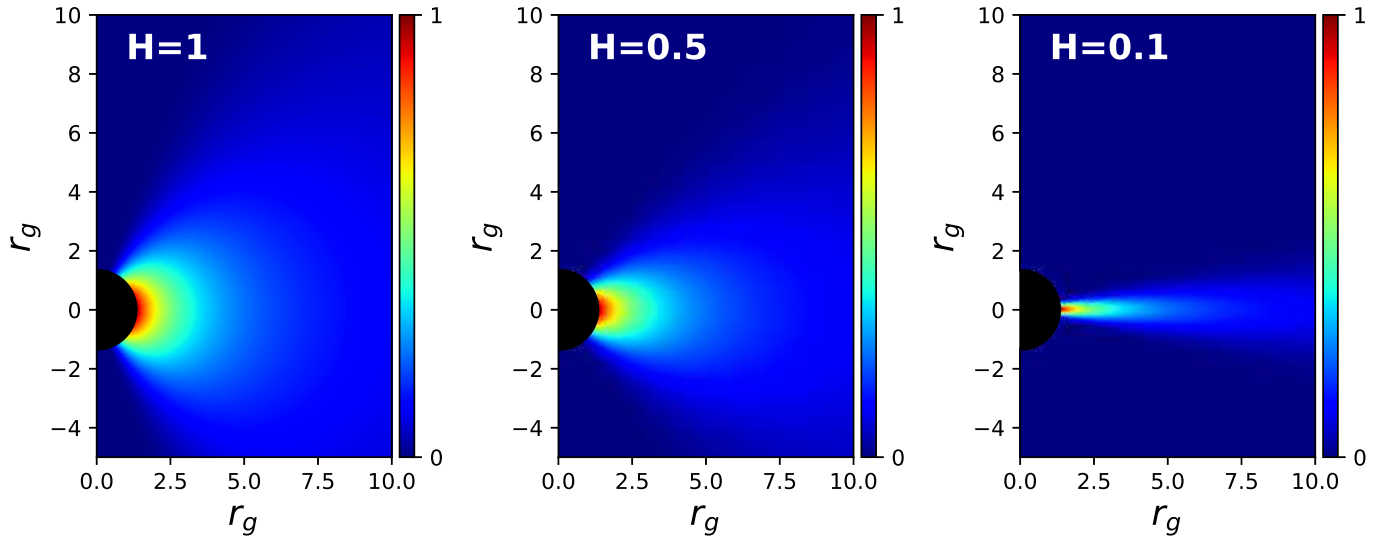}
    \caption{Spatial distribution of thermal electron, Equation (\ref{eq:p1}),  for $H=2$ (left), $H=1$ (middle), and $H=0.5$ (right).  The density is separately normalized and shown in linear scale, with red being larger and blue being small. A larger (or smaller) value of $H$ corresponds to a more spherical-like (or disk-like) geometry at the the innermost part of the flow. The black hole spin is $a=0.9375$.}  \label{fig:density}
  \end{center}
\end{figure*}

Radiative Inefficient Accretion flows (RIAF) around black holes at low accretion rates  is  the favored explanation for 
low-luminosity active galactic nuclei \citep[LLAGN;][]{ho09},  and black hole X-ray binaries  \citep[][]{nar96,esi97},  including the supermassive black hole at the
center of our Milky way galaxy, Sgr A$^{*}$ \citep[][]{nar95,yua03}. 
Very Long Baseline Interferometry (VLBI) observations from centimeter to millimeter wavelength show  that the  size of emission region in Sgr A$^{*}$ gradually decreases to its event horizon scale \citep[][]{kri99,lo99,bow06,doe08,fal09}. The innermost flow region of an RIAF is expected to be resolved by  millimeter/sub-millimeter VLBI observations, such as those by the  Event Horizon Telescope \citep[EHT;][]{fish14,doe08,ric15,lu16}.

While an RIAF is geometrically thick far away from the black hole,
in its innermost part, close to the black hole event horizon, the flow geometry is further affected by general relativistic effects.
 In the absence of a net magnetic flux, a
{\em quasi-spherical} geometry near the horizon is formed if the flow is everywhere sub-Keplerian, while a {\em disk-like} 
structure 
on the equatorial plane 
near the horizon is formed if the flow is super-Keplerian near the the innermost 
stable circular orbit \citep[][]{abr81}.
These two types of angular momentum distributions can result in different pressure distributions inside the flow: a outward  
monotonically decreasing pressure profile in former case, and a profile with a local pressure maximum inside the flow in the latter case
\citep[see Figure 2 and 4 of][]{nar97}.
Spectral models of X-ray binaries suggests a modest value of viscosity \citep[][]{nar96,esi97}, which guarantees that an RIAF is 
everywhere sub-Keplerian, and hence have a quasi-spherical geometry near the black hole.
The flow geometry  of the hot, radiatively inefficient flow has a roughly constant value $h/R\sim0.5$ \cite[][]{nar97,yua14}, where $R$ is the radius and $h$ is the vertical height of the flow. 

 The relationship between flow geometry and radiative efficiency
becomes more complicated when when magnetic fields are taken into account. 
As demonstrated in GRMHD simulations, the accumulated polar magnetic flux can compress the flow height near the horizon \citep[][]{mck12}. Moreover, the accumulated magnetic flux varies with the accretion environment, which can be simulated with different numerical setting such as Standard And Normal Distribution (SANE) or Magnetically Arrested Disk (MAD) \citep[e.g.][]{nar12, tch15}. For example, the flow can be vertically squeezed by the strong magnetic field accumulated around black hole with a geometry  $h/R\sim0.05$, while it is geometrically thick, $h/R\sim0.3-1$, at large distances \citep[][]{tch15}.

Many additional potential complications exist regarding the modeling of the emission region of Sgr A$^{*}$.  Deviations from the accretion disk morphology have been proposed by a number of authors \citep{fm00,mos13,mos14,cha15a,res17,med17}.  These typically invoke substantial emission from a putative relativistic jet or wind component, despite the lack of an obvious outflow structure at other wavelengths, motivated by similarities between Sgr A$^{*}$ and other LLAGN \citep{fm00}.  Even within the accretion flow paradigm, the black hole spin and disk angular momentum need not be aligned, leading to Lense-Thirring precession and associated features within the accretion flow \citep{dex13}.

 Here we limit our focus to the cases where the RIAF rotational axis is aligned with the black hole spin, and we focus on the flow structure. Based on the RIAF model first proposed by \citet[][]{bro06}, and a recent modification by \citet[][]{pu16a}, we aim to qualitatively investigate and discuss the observational consequences for different innermost accretion flow geometries. Such models are physically motivated by the RIAF structure of \citet{yua03}, and have been employed to estimate accretion flow properties from the increasing EHT observations of Sgr A$^{*}$ \citep[][]{bro09, bro11a, bro11b,bro16}, in which consistent results are obtained over many years,   providing physical understandings for  more complicated models.

The paper is organized as follows. We introduce the phenomenological model in \S \ref{sec:accmodel}, and present the results in \S \ref{sec:results}. Finally, the discussion and summary are given in \S \ref{sec:summary}.

\section{Innermost flow geometry of an RIAF: a Phenomenological Model }
\label{sec:accmodel}

\begin{figure*}
\begin{center}
\includegraphics[angle=0,width=\textwidth]{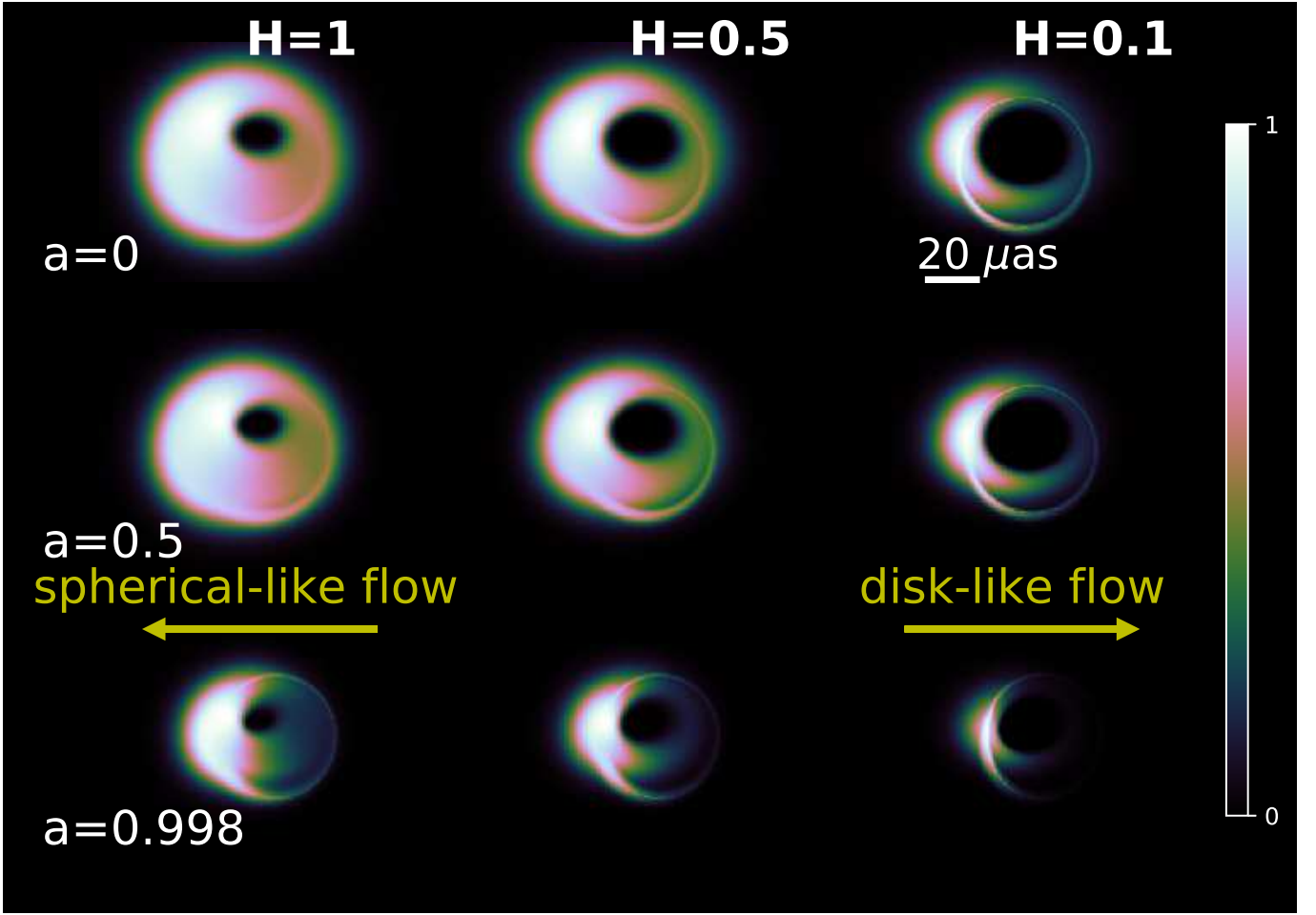}
\caption{Model Images of an Keplerian-rotating flow with different flow geometries at the innermost region, around a black hole with different  dimensionless spin parameter $a=0$ (top), $a=0.5$ (middle), and $a=0.998$ (bottom) at 230 GHz. A vertical spin axis is assumed.  
The flow with a smaller (or larger) value of $H$ is more disk-like (or spherical-like), as indicated by the arrows. The linear intensity scale (white is bright and green/black is faint) is normalized separately for each image. The inclination angle is $i=45^{\circ}$.  The electron number density and temperature are fixed for all panels of fixed spin parameter.}  \label{fig:ima_kep}
\end{center}
\end{figure*}

In this work, we focus on upcoming sub-mm VLBI observation of Sgr A$^{*}$ near millimeter wavelength $\sim$230 GHz (1.3 mm). 
At this frequency, the spectrum of Sgr A$^{*}$ transits from an inverted slope at lower frequency to a negative one, indicating the accretion flow environment is becoming transparent, the emission arises from a region very close to the black hole, and the shadow cast by the black hole can be observed \citep[][]{bar73,fal00}.
As the emission is dominated by thermal synchrotron emission (e.g.,  \citet[][]{oze00, mel00, mel01, yua03}; see also \citet[][]{mao17} for the effect of non-thermal electrons), we only consider the thermal synchrotron emission in this paper.
We consider that the spatial distributions of the electron temperature $T_{e}$ and thermal electron density $n_{e}$ are described by a hybrid combination of electrons
\begin{eqnarray}
n_{e}   &=& n^{0}_{e,\rm th}\;\,r^{-\alpha} e^{-\beta}\;, \label{eq:p1} \\
T_{e}             &=& T_{e}^{0}\;\,r^{-\gamma}\;, \label{eq:p2} 
\end{eqnarray}
where the radial dependence $\alpha=1.1$ and $\gamma=0.84$ are adopted from the vertically averaged density and temperature profiles found in \citet[][]{yua03},
and 
\begin{equation}
\beta=z^{2}/(2\sigma^{2})\;\;\;({\rm where}\;\; z\equiv r\cos\theta)\;.
\end{equation}
The parameter $\sigma$, therefore, controls the thickness of the accretion flow (in $z$-direction), and can be parameterized by
\begin{equation}
H\equiv\frac{\sigma}{x}\;\;\;({\rm where}\;\; x\equiv r\sin\theta)\;.
\end{equation}
The geometry of the flow is then controlled by the value of $H$, which is set to be unity in all previous works \citep[e.g.,][]{bro06, bro09, bro11a, bro11b,bro16,pu16a}. 

By introducing the parameter $H$, we can phenomenologically explore how  the innermost flow geometry, or the disk height of the accretion flow,  would affect the observed image and spectra.
A thicker (or thinner) flow geometry, and
 corresponding to a more ``spherical-like''  (or ``disk-like") flow, is represented by a larger (or smaller) value of $H$.
In the following we adopt three representative values of $H$= (1, 0.5, 0.1) to mimic flow  innermost geometries from a  spherical-like  to a  disk-like  geometries. 
For instance, the resulting density distribution for these three chosen values of $H$ around a fast-rotating black hole (with the dimensionless black hole spin parameter $a=0.9375$) are shown in Figure \ref{fig:density}. 

In the following, we compare and discuss the combined effects of the black hole spin, flow dynamics, and flow geometry. 
 We use \texttt{Odyssey} \cite[][]{pu16b} to perform the general relativistic radiative transfer computations\footnote{A polarized radiative transfer scheme is newly implemented into \texttt{Odyssey}, see the Appendix \ref{app}.}.
A angle-averaged emissivity  for thermal synchrotron emissivity is adopted for the computation of unpolarized emission \cite[][]{mah96}. The magnetic field strength $B$ is assumed to be in approximate equipartition with the ions, 
\begin{equation}
\frac{B^2}{8\pi}=\epsilon\,n_{e}\frac{m_{\rm p}c^2 r_{g}}{6r}\;,
\end{equation}
where $r_{g}=GM/^2$, $m_{\rm p}$ is the proton mass, and $\epsilon=0.1$, as considered in \citet[][]{bro11a, bro11b,bro16}.

\section{Results}
\label{sec:results}
We first compare the images of different flow geometries for the case of different black hole spins, with a mild inclination angle, $i=45^{\circ}$. 
 Considering an accretion flow which is Keplerian rotating outside the inner most stable circular orbit (ISCO), and infalling with a constant angular momentum inside ISCO , the computed black hole images at 230 GHz are shown in Figure \ref{fig:ima_kep}. The normalization of electron distribution $n^{0}_{e}$ ($\approx 10^{7}\;{\rm cm}^{-3}$) and temperature $T^{0}_{e}$ ($\approx 10^{11}\;{\rm K}$) reported in \citet[][]{bro06} is adopted.
 Located at $\sim8.3$ kpc and with a mass of $4.3\times10^{6} {\rm M}_{\odot}$ \citep[][]{ghe08,gil09a,gil09b}, the angular diameter of the photon ring (the collection of observed emission from the photon orbit(s) around a black hole, which marks the boundary of the black hole shadow) for Sgr A$^{*}$  is $\sim 50 \mu$as \cite[for a review, see e.g.][]{fal13}.
In general, due to the flow dynamics and rotation of spacetime, the flow at the approaching side and receding side transforms to be optically thin at different observational frequency. The optically thick part usually results in an image component with a crescent-like shape. 
At modest inclinations like the case here, the crescent surrounds a ``dim'' region with less emission close to the black hole spin axis.
A more disk-like flow with less material (see also figure \ref{fig:density}) results in a smaller size of the crescent component, and the ``dim" funnel  region becomes larger. 
In addition, as the size of the crescent component becomes smaller  for a high-spin black hole or a disk-like flow, the ``photon ring" surrounding the black hole shadow becomes more visible.
The black hole images for a similar inclination angle from post-processing numerical GRMHD simulations share the same feature \citep[e.g.,][]{mos09,dex10}.

\begin{figure}
\begin{center}
\includegraphics[angle=0,width=1.\columnwidth]{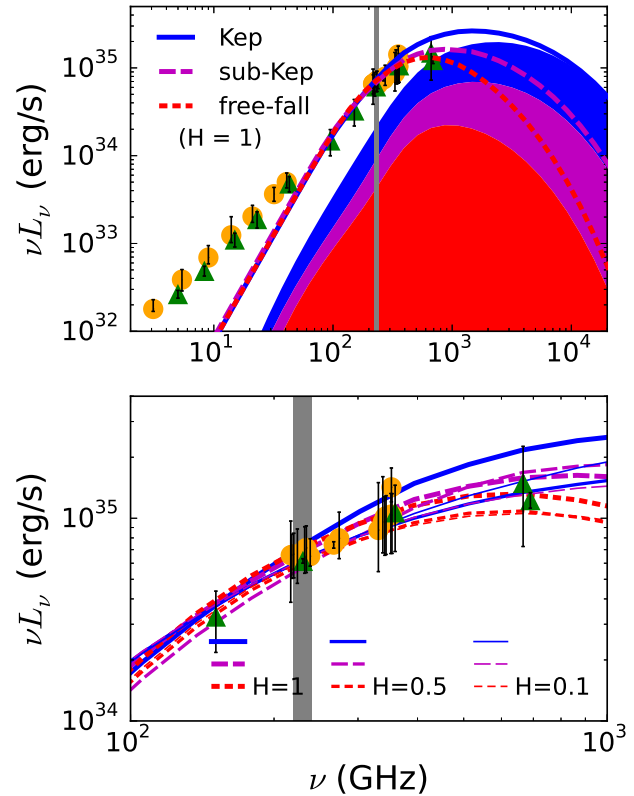}
\caption{{\em Top}: Model spectra of different accretion flow models around a rotating black hole with $a=0.9375$ and $H=1$: Keplerian flow (blue solid), sub-Keplerian flow (purple long dashed), and free-fall flow (red short dashed).  Observational data are from \citet[][and references therein, green triangles]{yua04}   and \citet[][yellow circles]{bow15} are overlapped. 
The inclination angle is $i=85^{\circ}$.  For reference, spectra of the emission from the region inside ISCO for Keplerian, sub-Keplerian, and free-fall flows are separately represented by the shaded blue, purple, and red regions.
{\em Bottom}: Zoom in plot of the spectrum near 230 GHz.  Cases of  $H=0.5$, and $0.1$ for all different flow dynamics are additionally shown in median-thickness and thin lines, respectively. The corresponding images at 230 GHz (indicated by the vertical grey area) are shown in Figure \ref{fig:ima_all}. See the text for more details.}  \label{fig:spec_all}
\end{center}
\end{figure}

\begin{figure*}
\begin{center}
\includegraphics[angle=0,width=\textwidth]{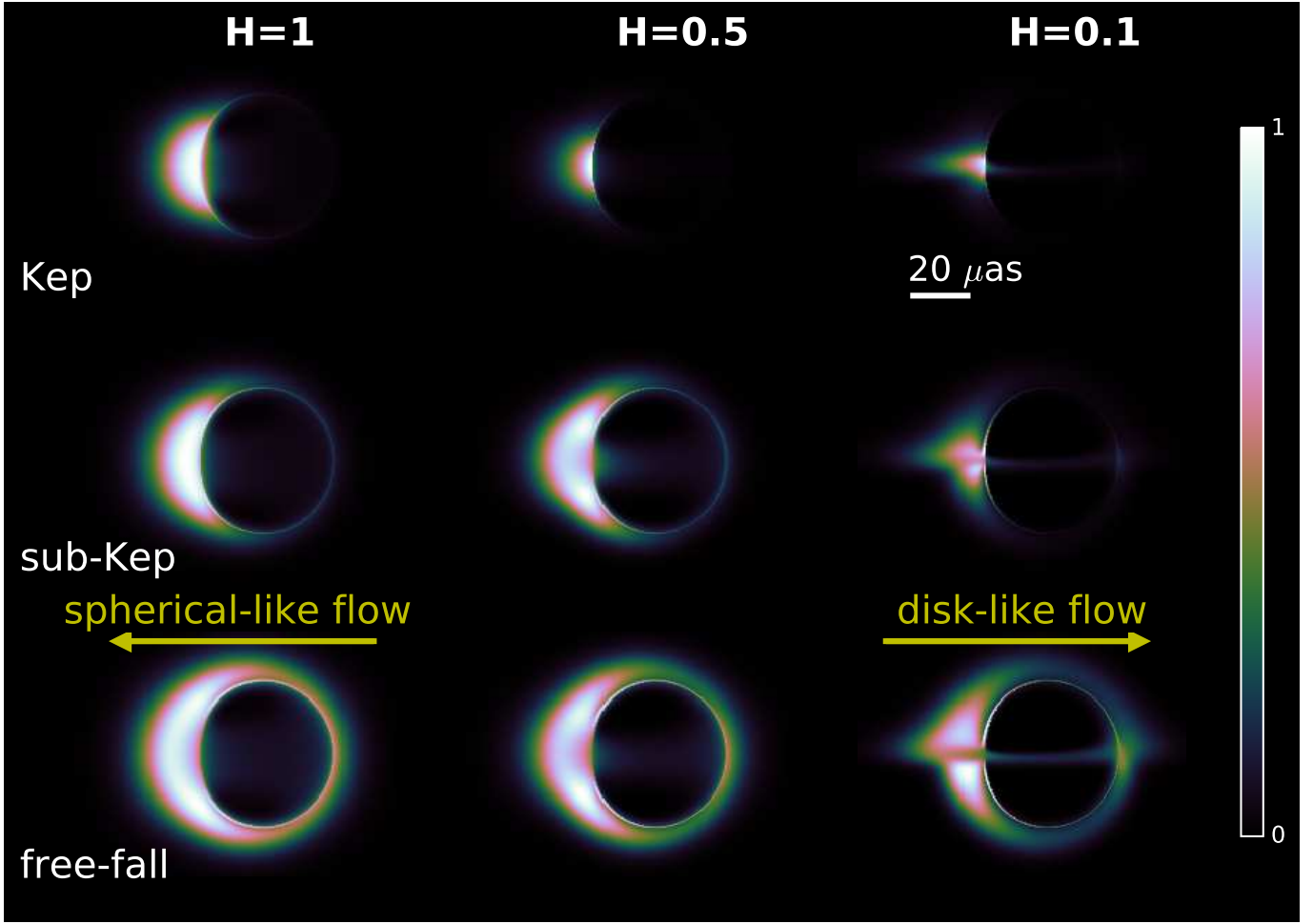}
\caption{Spectral fitted model Images of a fast rotating black hole with flows of different geometries of the innermost region, and different flow dynamics: Keplerian (top), sub-Keplerian (middle), and free-fall (bottom) at 230 GHz. The black hole spin is $a=0.9375$ and the inclination angle is $i=85^{\circ}$. A vertical spin axis is assumed. The flow with a smaller (or larger) value of $H$ is more disk-like (or spherical-like), as indicated by the arrows. The electron number density distribution for different values of $H$ can be referred to Figure \ref{fig:density}. The linear intensity scale (white is bright and green/black is faint) is normalized separately for each image.  The electron number density and temperature for each image are separately adjusted to generate similar observed flux ($\sim$ 3 Jy) at 230 GHz.}  \label{fig:ima_all}
\end{center}
\end{figure*}

While Keplerian flow dynamics is adopted when computing the images presented Figure \ref{fig:ima_kep},
and a sub-Keplerian flow dynamics is expected to be an important feature of an RIAF \citep[][]{ich77, nar94}, we next compare cases for different flow dynamics. A sub-Keplerian flow dynamical model can be constructed by a combination of a Keplerian flow and a free-fall flow which has zero angular momentum at infinity \citep[][]{pu16a}.

To gain more insight  into the combined effect for black hole spin, flow dynamics and flow geometry, we choose a high value of black hole spin, $a=0.9375$. In addition, a large inclination angle, $i=85^{\circ}$, is adopted to enhance the lensing effect for the flow on the equatorial plane. The electron density distribution is shown in Figure \ref{fig:density}.

The normalizations $n^{0}_{e,\rm th}$ ($\approx 10^{6-7}\;{\rm cm}^{-3}$)  and $T^{0}_{e}$ ($\approx 10^{11}\;{\rm K}$) of different flow dynamics and flow height are separately chosen to the values such that the thermal synchrotron emission of each case fits the observed flux of Sgr A$^{*}$ around 230 GHz ($\approx$3 Jy),  as shown in Figure \ref{fig:spec_all}. 
To generate a similar amount of luminosity observed at GHz frequency, the fitted spectra can be produced by a higher values of $T^{0}_{e}$ or  $n^{0}_{e,\rm th}$,  for a more sub-Keplerian flows to compensate for the weaker Doppler boosting (compared with the Keplerian case) at the approaching side, or for a more disk-like flow to compensate for the smaller emission region. As all these effects become significant when the accreting system become transparent, the differences between profiles become obvious when the accreting system becomes optically thin at  frequency  $\gtrsim$ hundred GHz. 

 The resulting spectra near 230 GHz for all cases are shown in  Figure \ref{fig:spec_all} (bottom panel). To highlight the emission contributed from different radius,  we overlapped (the shaded regions) the spectra resulting from flows inside ISCO\footnote{Recall that the photon ring of a rotating black hole is a combined contribution from unstable circular photon orbits. For $a=0.9375$, $r_{\rm ph}^{-}$($\simeq$3.94)$> r_{\rm ISCO}$($\simeq$2) $>r_{\rm ph}^{+}$($\simeq$1.44), where ``$-$'' denotes the outermost (retrograde) photon orbit and ``+'' denotes the innermost (prograde) photon orbit.} for the cases of $H=1$  in the top panel of Figure \ref{fig:spec_all}.
As expected,   more and more  emission  is contributed from regions closer to the black hole  as observational frequency increases.
Near 230 GHz, among all the different flow dynamics, the emission  from inside the ISCO for the free-fall flow is lowest (only $\approx 10\%$; the red shaded region), implying a more extended image due to the emission from outside ISCO, as will be shown in Figure \ref{fig:ima_all} later.  In contrast, for the Keplerian flow case, about  $30\%$ of the total flux are contributed from flows inside ISCO, implying a relative compact image size.

The computed model images for different flow dynamics and flow geometry at 230 GHz are shown in Figure \ref{fig:ima_all}. 
Unlike the modest inclination angle shown in Figure \ref{fig:ima_kep},  the crescent structure  and the photon ring are less overlapped.
As expected, a Keplerian flow (top panel) has a larger rotational velocity than other flow dynamics,  therefore the brightness contract between flow approaching/left side and the receding/right side is most obvious. For the same reason, the photon ring structure is most obvious for the free-fall flow. In contrast with the non-spinning black hole case \citep[][]{pu16a}, the spinning black hole images are asymmetric even for a free-fall accretion flow because of the rotation of the spacetime.

For all cases, the size of the crescent component at the approaching/left side decreases (correspondingly, the size of the dim region near the polar axis increases) when flow become more disk-like (see also Figure \ref{fig:ima_kep}).
The absence of material in the funnel region also leads to another important feature: the  innermost region of the flow behind the black hole  is seen via gravitational lensing. As the flow becomes more disk-like, the {\em shape} of the crescent structure at the approaching/left side appears more irregular.
This is because the crescent size in the horizontal  direction (perpendicular to the spin axis) is mainly contributed by
 the flow near the equatorial plane and therefore does not  significantly vary for different flow heights; while the crescent size in  the vertical direction (parallel to the spin axis) becomes more dominated by the emission from the flow behind the black hole as the flow become more disk-like. 

Furthermore, the {\em internal brightness distribution} within the crescent structure becomes less smooth due to  the combined contribution from flows in front of or beside the black hole, and from flows behind the black hole.   
With similar features seen in the disk-like flow in
a modest inclination cases shown Figure \ref{fig:ima_kep}, a bright crescent component plus a bright photon ring component, a dimmer region of the crescent structure could exist between its brighter region and the photon ring in more edge on cases, as shown in the rightmost panels of Figure \ref{fig:ima_all}. 
 
\section{Discussion and Summary}
\label{sec:summary}

\begin{figure*}
\begin{center}
\includegraphics[angle=0,width=1.\textwidth]{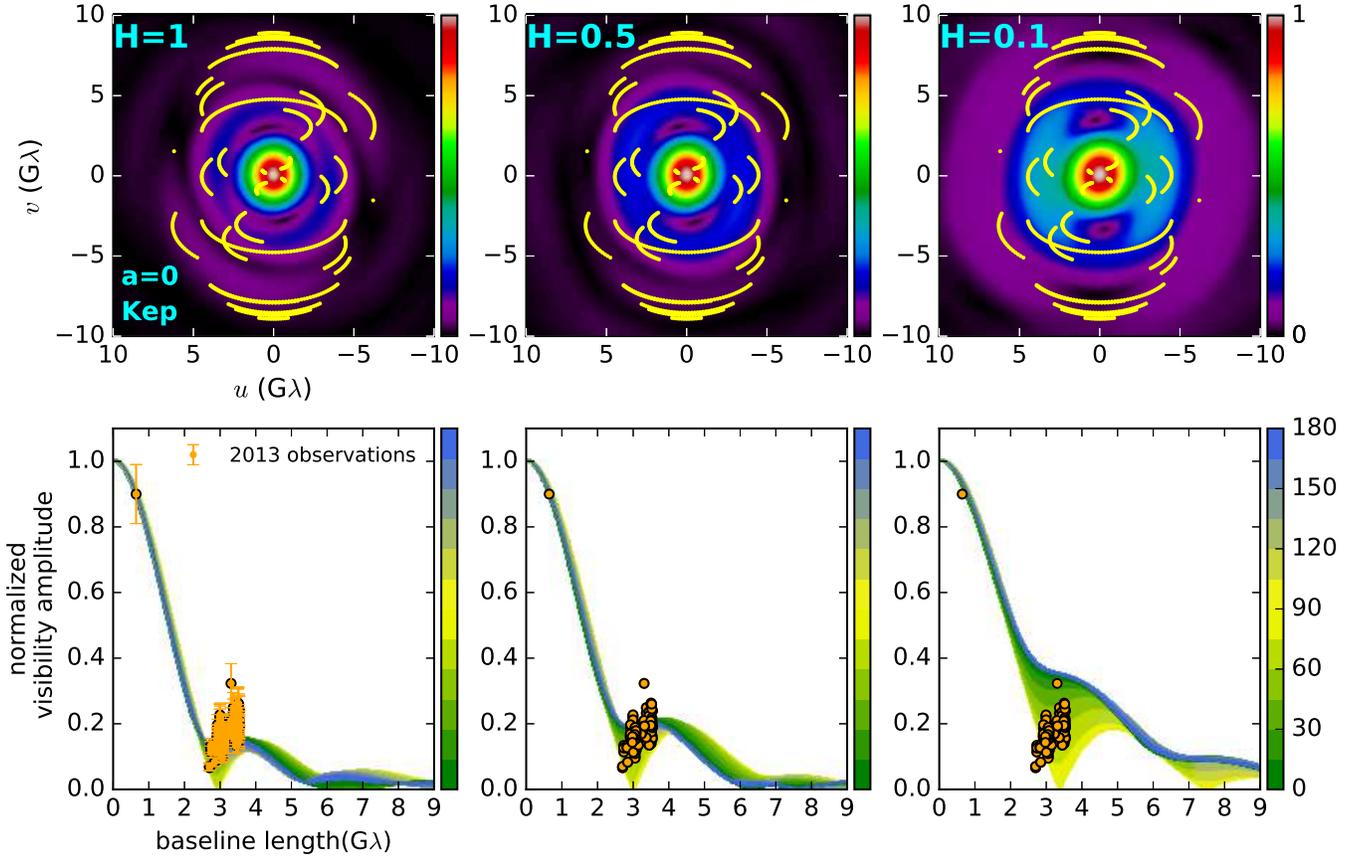}
\caption{Model visibility amplitude map of the model images for the case of $a=0$ in figure \ref{fig:ima_kep} (top), and their values as function of the baseline length (bottom).
The visibility amplitudes is normalized by the total flux density for each image separately. The linear intensity scale (red is bright and blue is faint) is normalized separately for each image.
An example $u-v$ coverage of the full array EHT is overlapped with the visibility maps on the top with the yellow lines, assuming a zero position angle for the source. The color of the bottom panel represents the profile along a slice passing through $(u,v)=(0,0)$, with a clock-wise angle (in degree) from $v=0$ and $u>0$ on the map. The de-blurred normalized visibility of the 2013 EHT observations \citep[][]{joh15} are overlapped with the amplitude-baseline length plots on the bottom. Error bars of the data are not shown for  the case of $H=0.5$ and $H=0.1$. A vanishing position angle is assumed.}  \label{fig:vis_kep0}
\end{center}
\end{figure*}

\begin{figure*}
\begin{center}
\includegraphics[angle=0,width=\textwidth]{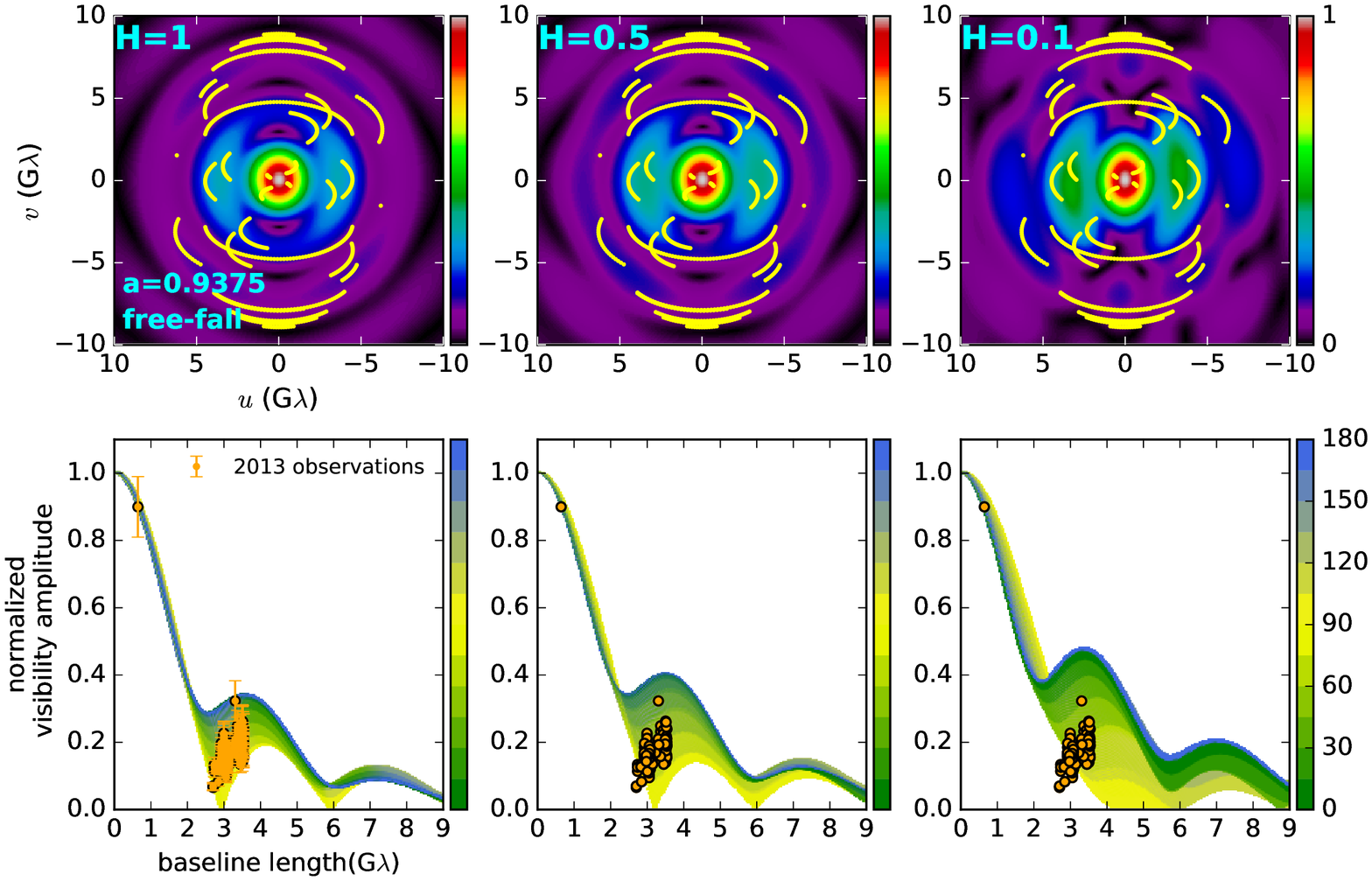}
\caption{Model visibility amplitude map of the free-fall model images shown in Fig \ref{fig:ima_all} (top), with their values versus baseline length (bottom).  For  more disk-like accretions, the first minimum of the visibility amplitude move further away and the variations of visibility amplitude becomes larger. See \ref{sec:dis_vis} for more discussions.}  \label{fig:vis_ff}
\end{center}
\end{figure*}

 \subsection{Visibilities}\label{sec:dis_vis}
Can we distinguish the different image features mentioned above from VLBI observations?
The visibility, a observable for VLBI,  is related to the Fourier transform of the model image \citep[e.g.,][]{tho17}. 
As discussed below, there are several distinguishable visibility features for different flow properties at their innermost region.

In Figure \ref{fig:vis_kep0}, we present the normalized model visibility amplitude map (top panel) and their profile along base line length (bottom panel) for one of  the spin ($a=0$ and $i=45^{\circ}$) in Figure \ref{fig:ima_kep}. 
This is close to the reported best fit of Sgr A$^{*}$, $a=0$ and $i=68^{\circ}$,  obtained from EHT observations \citep[][]{bro09,bro11a};  note also other estimated black hole spin in \citet[][$a=0.9$]{mos09}, \citet[][$a\lesssim0.9$]{hua09a}, and \citet[][$a=0.5$]{shc12}.
For reference, the $u-v$ coverage of future EHT observations\footnote{See the VLBI Reconstruction Dataset webpage: \url{http://vlbiimaging.csail.mit.edu/training_02}} are overlapped with the visibility maps (top panel), in which a black hole spin axis parallel to vertical direction (i.e., a vanishing position angle) is assumed. 
The normalized visibility amplitude versus baseline length (bottom panel) along different slices through $(u, v)=(0, 0)$ with a clockwise-rotating angle from $v=0$ and $u>0$ are plotted by different color.
For reference, the de-blurred normalized visibility of the 2013 EHT observation with CARMA at California, SMT at Arizona, and SMA/JCMT at Hawaii \citep[][]{joh15} are overlapped with the plot.

As the flow becomes more disk-like, the decrease of the crescent structure in  the vertical direction (which is more obvious than that in horizontal direction), as shown in Figure \ref{fig:ima_kep}, results in an increase in size within the visibility map along $u\sim0$  (see the yellow profiles shown in the bottom panel), recalling that model image and visibility map are Fourier pairs. 
The first minimum of the visibility amplitude along baseline length ($\sim$3G$\lambda$) therefore move farther away.
Intriguingly, the the minimum of the visibility amplitude is due to the dim funnel near the pole region in each model images, instead of the photon ring around the black hole shadow.
While current EHT observations provide data within 4G$\lambda$ baseline length and the baseline length $\sim$3G$\lambda$ provides important observational constraints for RIAF parameters \citep[][]{fish09}, the visibility profile and the second minima at longer baselines from future  EHT observations will provide further information of the flow structure. However, note that at long baselines a single mm-submm VLBI observations of Sgr A$^{*}$ suffer from both diffractive and refractive scattering in the interstellar medium \citep[][]{gwi14, jg15,jn16}.

Further visibility features can be revealed by examined a more edge-on case, as presented in Figure \ref{fig:ima_all}.
The corresponding model visibilities for the free-fall images (bottom panel of  Figure \ref{fig:ima_all}) are shown in  Figure \ref{fig:vis_ff}. 
It is important to note that the variations of visibility amplitude at a given baseline length become larger for disk-like flow, as a result of a more complicated, mixed brightness distribution within the crescent structure. This feature is also shown in Figure \ref{fig:vis_kep0}.
The first minimum of visibility amplitude along baseline length is determined by the largest dimension of the crescent component (compare the yellow profiles of the bottom panel),  and is usually dominated by the flow height. 
Other model images shown in Figure \ref{fig:ima_all}  have a visibility versus baseline profile whose first minimum is located farther away from the observation data and hence not shown.

\subsection{Polarization} \label{sec:dis_pol}
\begin{figure*}
\begin{center}
\includegraphics[angle=0,width=\textwidth]{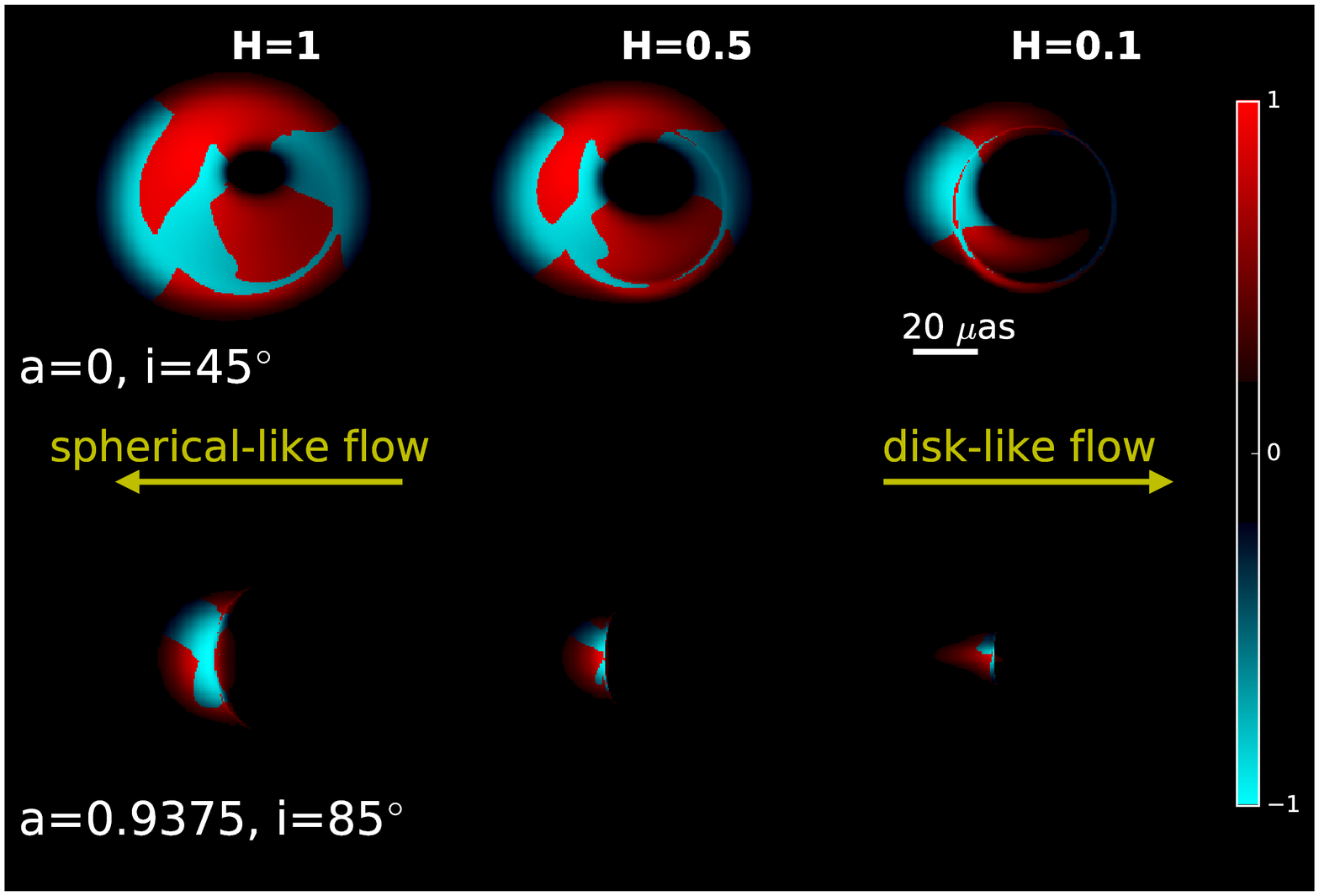}
\caption{ Polarization maps of Keplerian flows shown in top panels of Fig \ref{fig:ima_kep} (top) and Fig \ref{fig:ima_all} (bottom), assuming the local magnetic field is purely toroidal. The vertical polarized emission is in red  (positive Stoke Q) and the horizontal polarized emission  (negative Stoke Q) is in cyan . The pixel brightness is  linearly scaled by the total flux,  as indicated by the colorbar. For more disk-like accretions, lensed emission from the flow behind the black hole leads to a more complicated polarization distribution within the crescent structure. See Section \ref{sec:dis_pol} for discussions.}  \label{fig:kep_pol}
\end{center}
\end{figure*}

The detailed polarization property of the flow is related to several factors, such as the local magnetic field configuration, the flow motion, the opacity of the flow, the polarization degree \citep[e.g. ][]{pet83,pan16, dex16}, the Faraday effect \citep{dex16, mos18},  the additional emission contributed from non-thermal electrons \citep{bro06}, and the combination thereof.
While a comprehensive study including all of these effects is beyond the scope of current paper, 
here we qualitatively explore how the polarzation properties would depend on accretion flow geometry.

The magnetic field configuration is expected to be coupled with the flow motion; for simplicity, here we focus on the the Keplerian flows (top panels of Figures \ref{fig:ima_kep} and \ref{fig:ima_all}) and assume the magnetic fields are purely toroidal.  For other flow dynamics, the field configuration can be more complicated.

We ignored the contribution of the Stokes $V$ and all Faraday terms
\citep[note the circular polarization fraction is low ($<1 \%$) for Sgr A$^{*}$ at $\gtrsim$ 100 GHz,][]{mun12}. The thermal synchrotron polarization fraction compared with the angle-averaged emission is estimated by the formula in \citet[][]{pet83}, and the contribution of the Stokes $V$ and all Faraday terms are ignored
(see the Appendix \ref{app} for a description of the polarized general relativistic transfer scheme).

The vertical (positive Stokes $Q$; in red) and horizontal (negative Stokes $Q$; in cyan) polarized emission on the sky plane is presented in Figure \ref{fig:kep_pol}.  
Without in situ Faraday rotation, the polarization direction is roughly perpendicular the field direction (with the lensing) in the optically thin regime. For a mildly inclined flow, shown in the top panel, the positive Stokes $Q$ regions therefore roughly occupies the upper and lower part of the crescent structure \citet[see also][for a similar result]{bro01}, while the details of the boundary between the red and cyan regions depend on the several factors mentioned previously.
For a more edged on case shown in the bottom panel, the the positive Stokes $Q$ additionally appears near the mid plane of the image.
As the accretion flow geometry becomes more disk-like, additional emission is contributed from the far side behind the black hole due to gravitational lensing. 
This implies that how the polarization dominance varies with increasing image resolution \citep[][]{joh15, gol17} also depends on the accretion geometry.

\subsection{Summary}
Different potential innermost-flow geometries and dynamics of RIAF models produce characteristic qualitative effects in the resulting images that are apparent in the brightness distribution of the crescent structure and its polarization.  These lead to qualitative modifications of the visibility amplitudes at intermediate and long baselines for current and future EHT observations.  The fine details of the crescent structure provide physical interpretations and possible new elements for the geometric crescent models \citep[e.g.,][]{kam13, ben16}, helping to produce high fidelity, phenomenological model images for interpreting EHT observations.

A number of potential additional systematic uncertainties exists.  In addition to modifications of the geometry emission region (e.g., jets, tilted disks, etc.), modifications of the underlying electron population beyond a thermal bump, the intrinsic radial structure of the emission component considered here may differ.  Addressing the full set of flow structural parameters must await a parameter-estimation study and is beyond the scope of this work.  However, we note that even substantial variations in the radial dependence of the electron number density and temperature in Equations (\ref{eq:p1}) and (\ref{eq:p2}) do not modify the qualitative conclusions reached here: the dynamics and geometry of the innermost portions of an accretion flow in Sgr A$^*$ will be discernible by EHT observations.

\acknowledgments
\label{ack}
We are grateful to Hotaka Shiokawa for stimulating discussions which help to initiate and improve this work.
We also thank Michael Johnson for providing the observation data, Ziri Younsi, Jason Dexter, and the anonymous referee for helpful suggestions and comments. A.E.B. receives financial support from the Perimeter Institute for Theoretical Physics and the Natural Sciences and Engineering Research Council of Canada through a Discovery Grant. Research at Perimeter Institute is supported by the Government of Canada through Industry Canada and by the Province of Ontario through the Ministry of Research and Innovation. This research has made use of NASA's Astrophysics Data System. 
 \software{\texttt{Odyssey}\citep[][]{pu16b}}

\appendix
\section{Polarized General Relativistic Radiative transfer Scheme}\label{app}
\begin{figure}
\begin{center}
\includegraphics[angle=0,width=0.5\textwidth]{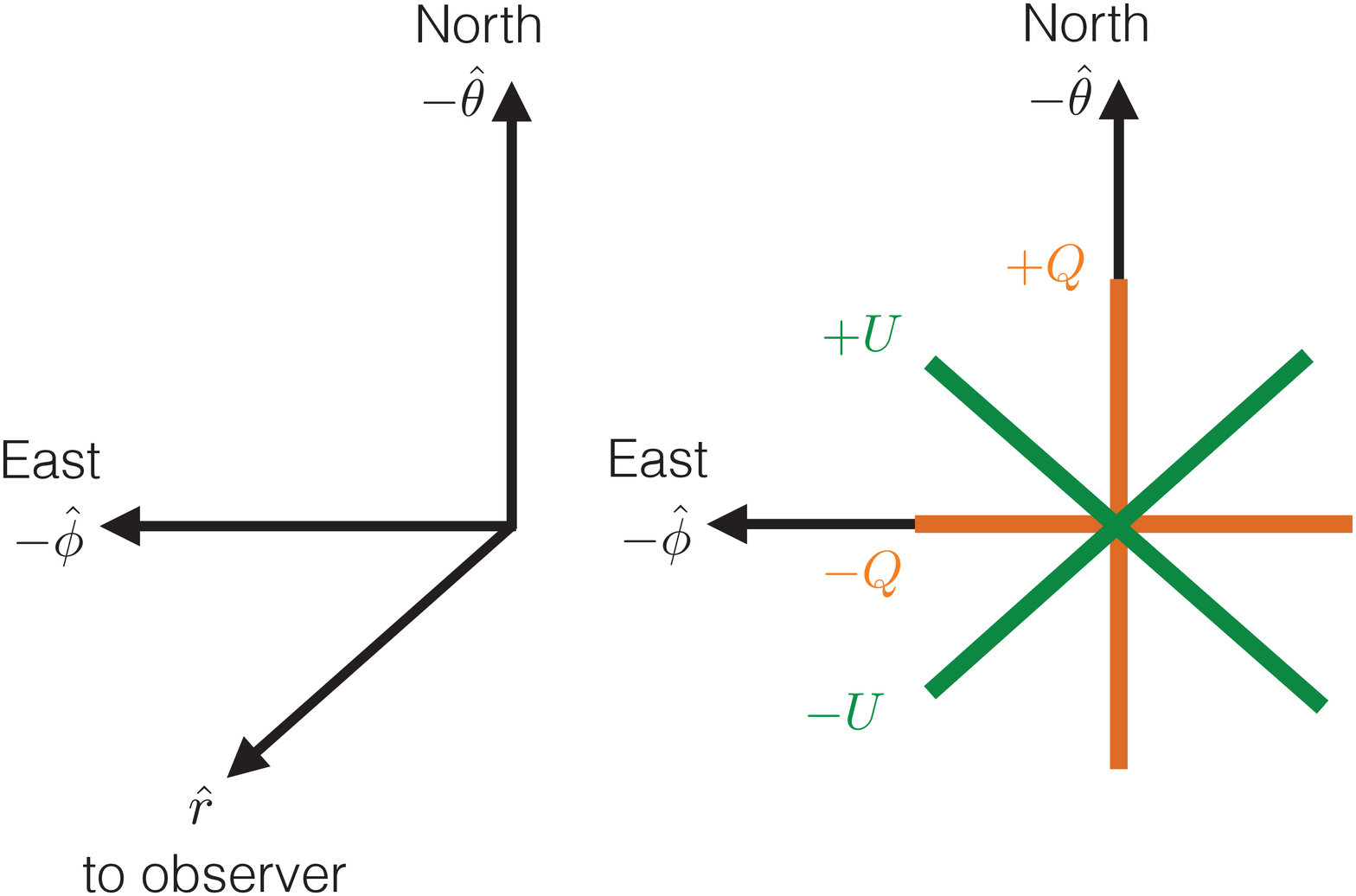}
\caption{{\em Left}: Polarization basis setup in the Boyer-Lindquist coordinate. {\em Right}: The electric vector position angle (EVPA) orientation for Stokes $Q$ and $U$, as seen in the observer's sky plane. The EVPA is defined by $\frac{1}{2}\tan^{-1}(U/Q)$, measured from East of North, following the IAU/IEEE definition.}  \label{fig:pol}
\end{center}
\end{figure}

Here we introduce the polarized general relativistic radiative transfer scheme  implemented into \texttt{Odyssey} for synchrotron radiation. 

The computation is performed in the Boyer-Lindquist coordinate, and we choose the polarization basis in observer's sky frame along with the $-\hat{\theta}$ (North) direction and $-\hat{\phi}$ (East) direction. Such
choice is consistent with the IAU/IEEE definition \citep{ham96}, as illustrated in Figure \ref{fig:pol} (see also figure 1 of \citet{ham96}). With the definition, the positive stoke parameter $Q$ is therefore measured according to the north direction. The polarized radiative transfer is performed by two steps, as described below.

We first trace the ray from observers' sky frame {\em backward} in time (from the observer to the source), by solving  
$(dt/d\lambda,\; dr/d\lambda,\; d\theta/d\lambda,\; d\phi/d\lambda,\; dP_{r}/d\lambda,\; dP_{\theta}/d\lambda)$ with Runge-Kutta method, as described in \citet{pu16b}.  Here $x^{\alpha}=(t, r, \theta, \phi)$ is the location of the photon and $P^{\alpha}$ is the four-momentum of the photon, and $\lambda$ is the affine parameter. At this step, the (unpolarized)  Lorentz invariant intensity $\mathcal{I}_{\rm unpol}\equiv I_{\rm unpol}/\nu^{3}$ and the optical depth $\tau$  can be computed by \citep{you12}
\begin{equation}
\frac{d\tau}{d\lambda}=g^{-1}\alpha_{I}\;,
\end{equation}
\begin{equation}\label{eq:I_backward}
\frac{d\mathcal{I}_{\rm unpol}}{d\lambda}=g^{-1} \frac{j_{I}}{\nu^{3}} {\rm exp}(-\tau)\;,
\end{equation}
where $\alpha_{I}$, $j_{I}$, $\nu$ are the absorption coefficient, emissivity coefficient, and the frequency in the fluid's comoving frame, respectively, and 
\begin{equation}
g\equiv\frac{\nu_{\rm obs}}{\nu}
\end{equation}
 is the frequency shift between the distant observer's frame ($\nu_{\rm obs}$) and the fluid's comoving frame ($\nu$). 
 
Next, once the ray leave the region of interest or enter a highly optically thick region, the direction of time is inverse in the differential equations, so the polarized radiative transfer equations are integrated {\em forward} in time (from the source to the observer). During this time-forward integration, in additional to the above six equations, four differential equations for the Lorentz invariant form of all four Stoke parameters $(\mathcal{I}_{\rm pol}, \mathcal{Q}, \mathcal {U}, \mathcal{V})\equiv$$(I_{\rm pol}/\nu^{3}, Q/\nu^{3}, U/\nu^{3}, V/\nu^{3})$ are also solved simultaneously along the ray (see equation (\ref{IQUV_forward}) later). We follow the polarized radiative transfer algorithm described in \cite{dex16} (see also \citet{hua09b, shc11, mos18}), in which the Walker-Penrose constant \citep{wal70} is used for tracing the rotation of polarization basis as the ray travel in the curved spacetime \citep{con80}, then a rotation matrix $\mathcal{R}(\chi)$ is applied for aligning the projected magnetic field in the plasma frame to one of the stoke $Q$ directions. The differential equations
 $(d\mathcal{I}_{\rm pol}/d\lambda,\; d\mathcal{Q}/d\lambda,\; d\mathcal{U}/d\lambda,\; d\mathcal{V}/d\lambda)$
  therefore have the form
\begin{equation}\label{IQUV_forward}
\frac{d}{d\lambda}
 \left(
\begin{array}{c}
\mathcal{I}_{\rm pol}\\
\mathcal{Q}\\
\mathcal{U}\\
\mathcal{V}
\end{array}     
\right)
=g^{-1}
\left[\mathcal{R}(\chi)
\left(
\begin{array}{c}
j_{I}\\
j_{Q}\\
j_{U}\\
j_{V}
\end{array}
\right)
-
\mathcal{R}(\chi)
\left(
\begin{array}{cccc}
\alpha_{I}  & \alpha_{Q} & \alpha_{U}             & \alpha_{V} \\
\alpha_{Q} & \alpha_{I} & \rho_{V}   & -\rho_{U}\\
\alpha_{U} & -\rho_{V}   & \alpha_{I} & \rho_{Q}\\
\alpha_{V} & \rho_{U}              & -\rho_{Q} &\alpha_{I}
\end{array}  
\right)  
\mathcal{R}(-\chi) 
 \left( 
\begin{array}{c}
\mathcal{I}_{\rm pol}\\
\mathcal{Q}\\
\mathcal{U}\\
\mathcal{V}
\end{array}     
\right)
\right]\;,
\end{equation}
with $j_{(I, Q, U, V)}$ and $\alpha_{(I, Q, U, V)}$ are respectively the emissivity and absorption coefficients associated with the Stoke parameters, and $\rho_{(Q, U)}$ and $\rho_{V}$ are respectively the Faraday conversion and rotation coefficients.

In general, the integrated intensity $I_{\rm pol}\neq I_{\rm unpol}$, and for a arbitrarily polarization radiation, $I_{\rm pol}\geqslant Q^{2} + U^{2} +V^{2}$. By aligning the projected magnetic field to $-Q$ direction,  we have $j_{Q}>0$, and $j_{U}=\alpha_{U}=\rho_{U}=0$ in equation (\ref{IQUV_forward}) by symmetry. 
The implementation of accurate polarized synchrotron emissivities and absorptivities such as those in \citet[][]{pan16} and \citet{dex16} is a straightforward future work.

\label{reference}
\end{document}